\documentclass[
aps,
prl,
reprint,
superscriptaddress,
amsmath,
amssymb,
longbibliography,
]{revtex4-1}

\usepackage{graphicx}
\usepackage{dcolumn}
\usepackage{bm}
\usepackage{mhchem}
\usepackage{natbib,hyperref}
\hypersetup{colorlinks=true,linkcolor=blue,citecolor=blue,urlcolor=blue}
\usepackage{color}
\usepackage{ulem}
\usepackage{float}

\renewcommand{\vec}[1]{\mathbf{#1}}
\renewcommand{\emph}[1]{\textit{#1}}

\begin{document}


\title{Near Unity Optical Spin Polarization in GaSe Nanoslabs}
\title{Optical Spin Polarization Dynamics in GaSe Nanoslabs}
\author{Yanhao Tang}
\author{Wei Xie}
\affiliation{Department of Physics and Astronomy, Michigan State University, East Lansing, MI 48824, USA}
\author{Krishna C. Mandal}
\affiliation{Department of Electrical Engineering, University of South Carolina, Columbia, SC 29208, USA}
\author{John A. McGuire}
\author{C. W. Lai}
\email{cwlai@msu.edu}
\affiliation{Department of Physics and Astronomy, Michigan State University, East Lansing, MI 48824, USA}


\begin{abstract}
{We report nearly complete preservation of ``spin memory" between optical absorption and photoluminescence (PL) in nanometer slabs of GaSe pumped with up to 0.2 eV excess energy. At cryogenic temperatures, the initial degree of circular polarization ($\rho_0$) of PL approaches unity, with the major fraction of the spin polarization decaying with a time constant $>$500 ps in sub-100-nm GaSe nanoslabs. Even at room temperature, $\rho_0$ as large as 0.7 is observed, while pumping 1 eV above the band edge yields $\rho_0$ = 0.15. Angular momentum preservation for both electrons and holes is due to the separation of the non-degenerate conduction and valence bands from other bands. In contrast to valley polarization in atomically thin transition metal dichalcogenides, here optical spin polarization is preserved in nanoslabs of 100 layers or more of GaSe.} 
\end{abstract}


\maketitle

Solid-state systems exhibiting high spin polarization and long spin relaxation time are desirable for spintronic applications. Various semiconductors have been studied for creation of long-lived non-equilibrium spin populations and coherences \cite{dyakonov1984,dyakonov2008,pikus1984,wu2010,zutic2004,awschalom2013}. The most extensively studied system is gallium arsenide (GaAs). However, optically pumped electron spin polarization in bulk GaAs is limited to 1/2, while the maximal degree of circular polarization of photoluminescence is 1/4 \cite{dyakonov1984,dyakonov2008}, owing to the degenerate heavy- and light-hole valence bands and sub-ps hole spin relaxation. Doping \cite{kikkawa1997} or quantum confinement \cite{ohno1999,winkler2003} has been used to quench \emph{electron} spin relaxation. Unity electron spin polarization can be achieved in heterostructures where heavy- and light-hole energy degeneracy is lifted by quantum confinement or strain \cite{kohl1991,amand1994,pfalz2005}. In analogy to spin polarization, valley polarization has been demonstrated in monolayer transition metal dichalcogenides (TMDs) with potential applications exploiting both spin and valley degrees of freedom. In monolayer TMDs inversion symmetry is broken, and a direct gap emerges at the corners ($K$ points) of the Brillouin zone, enabling valley-dependent inter-band transitions under circularly polarized optical excitation \cite{yao2008,cao2012,zeng2012}. Furthermore, the substantial spin-splitting of valence bands at the band edges due to spin-orbit interaction originating from the $d$ orbitals of TM ions has led to recent reports of long hole spin and valley lifetimes \cite{mak2012}, valley exciton polarization and coherence \cite{xu2014,zhu2014}, circularly polarized electroluminescence \cite{zhang2014}, and valley Hall effect \cite{mak2014}. Indeed, circularly polarized PL was observed in single- and bi-layer TMDs \cite{cao2012,zeng2012,mak2012,xu2014,zhu2014} with steady-state near-resonant circularly polarized excitation. However, time- and polarization-resolved PL measurements suggest that, at least in \ce{MoS2}, circularly polarized PL may result from sub-10-ps recombination and valley (spin) lifetimes rather than an intrinsically long-lived valley or hole spin polarization \cite{cao2012,lagarde2014,glazov2014}. Additionally, emission at the direct gap becomes dominant only at the monolayer level \cite{splendiani2010,mak2010,tonndorf2013}.

\begin{figure}[htbp]
\includegraphics[width=0.45 \textwidth]{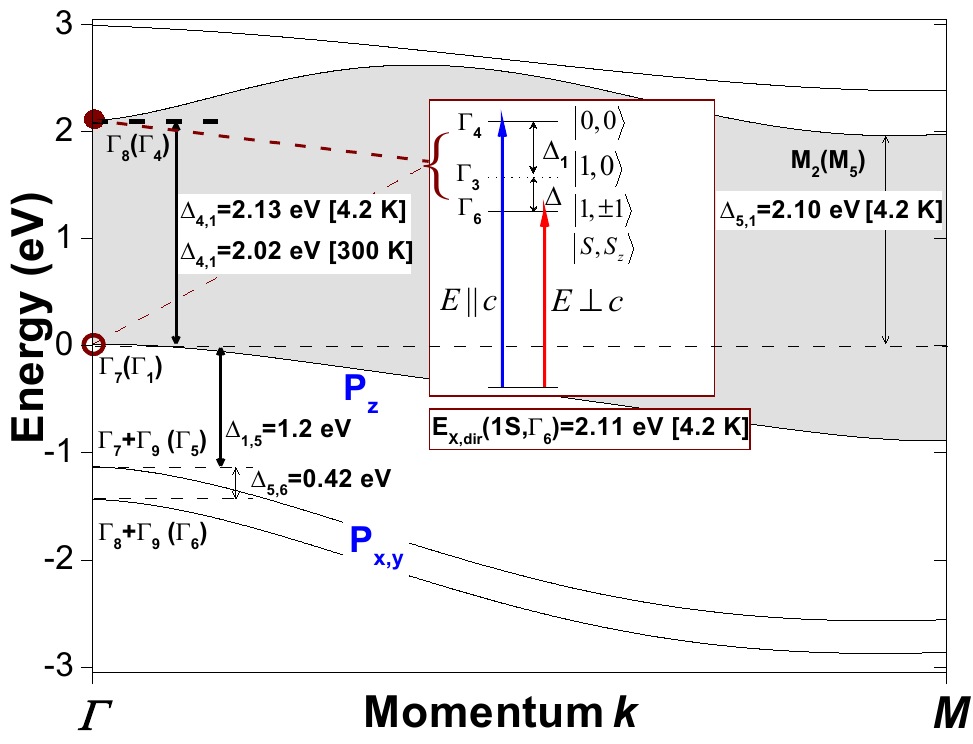}
\caption{\label{fig:band}(color online) \textbf{Band structure and optical selection rules.} Sketch of the band structure of $\epsilon$-GaSe at the $\Gamma$ point and the representations to which the states at the $\Gamma$-point belong with (without) spin-orbit interaction. The inset shows the excitons and optical selection rules.}
\end{figure}

In this study, we demonstrate GaSe as a material for generating and preserving a high degree of spin polarization under nonresonant optical pumping, i.e., the energy difference between the pump and PL spectral peak (excess energy) is more than 0.1 eV. The unique bandstructure (Fig.~\ref{fig:band}) of the group-III monochalcogenides removes degeneracies between orbital states and thereby allows generation and preservation of a high degree of spin polarization \cite{gamarts1977,ivchenko1977}. We investigate the photoluminescence of excitons in sub-100-nm to 1000-nm thick GaSe slabs (nanoslabs) under nonresonant circularly polarized optical excitation. Polarized time-dependent PL in GaSe reveals a high initial degree of circular polarization $\rho_0>0.9$ when excited with excess energy from about 0.1 to 0.2 eV. High $\rho_0$ is indicative of spin preservation of optically active carriers during the optical absorption and subsequent cooling process. Unlike TMDs, GaSe has a quasi-direct gap in the bulk form \cite{mooser1973,capozzi1993}, making it a versatile material in which strong emission occurs and light-matter coupling can be controlled from bulk to nanoscale thicknesses.

The crystal structure of $\epsilon$-GaSe shows an $ABA$ stacking of the individual layers, and belongs to space group $D_{3h}^{1}/P\bar{6}m2$ (\#187) and point group $D_{3h}$. An individual layer consists of four planes of \ce{Se-Ga-Ga-Se}, with the \ce{Ga-Ga} bonds normal to the layer plane and arranged on a hexagonal lattice. The Se anions are located in the eclipsed conformation when viewed along the $c$-axis. The physical properties of \ce{GaSe} are largely determined by those of a single layer due to the high electron density within the layer and weak interlayer interaction \cite{mooser1973,schluter1976}. The Se-Ga bond tilts 29$^\circ$ out of the layer plane so that Se 4$p_{x,y}$ electrons experience much greater Coulomb attraction with \ce{Ga} cations than Se 4$p_z$ electrons do. This results in a large negative crystal field ($\Delta_{cr} \approx$ -1.4 eV) compared to wurtzite-type materials ($\Delta_{cr} \sim$ +0.05 eV) so that the Se 4$p_{x,y}$ states lie 1.2 and 1.6 eV below the Se 4$p_z$ state (Fig. \ref{fig:band}). Consequently, near the $\Gamma$ point ($\vec{k}=0$) only light with $\vec{E} \parallel c$ can excite dipole transitions between the $p_z$-like uppermost valence band and $s$-like lowest conduction band. Experimentally, the absorption coefficient for $\vec{E} \perp c$ is about $10^3$ cm$^{-1}$, a factor of 30 smaller than for $\vec{E} \parallel c$ \cite{le-toullec1977}. The transition for $\vec{E} \perp c$ is weakly dipole-allowed owing to band mixing induced by strong spin-orbit interaction ($\Delta_{so} \, \approx$ 0.44 eV) and weak interlayer coupling. The optical pumping and selection rules near $k=0$ are best understood in an exciton (two-particle) picture (binding energy $\sim$ 20-30 meV) \cite{mooser1973,tang2014}. The essential feature is that the lowest states correspond to total exciton spin $S=1$ with $z$-component $S_z=\pm1$ and can be excited by light with wave vector $\vec{k} \parallel c$ ($\vec{E} \perp c$). We have exploited these selection rules to investigate the spin dynamics of GaSe under nonresonant circularly polarized optical excitation as a function of slab thickness $d_L$.

\paragraph{Experiment.} Nanometer thick GaSe crystals are mechanically exfoliated from a Bridgman-grown crystal with unintentionally \emph{p}-doped concentration $10^{14}-10^{15}$ cm$^{-3}$ \cite{mandal2008a} and deposited onto a silicon substrate with a 90 nm \ce{SiO2} layer. Sample thickness is measured by atomic force microscopy. Samples are mounted in vacuum on a copper cold finger attached to an optical liquid helium flow cryostat for all experiments. GaSe nanoslabs are optically excited by 2 ps laser pulses from a synchronously pumped optical parametric oscillator ($\lambda_p \sim$ 560--595 nm, $E_p \sim$ 2.21 -- 2.08 eV) or by second-harmonic pulses from a Ti:sapphire oscillator ($\lambda_p \sim$ 410 nm, $E_p \sim$ 3.0 eV). The laser beam is focused through a microscope objective (numerical aperture N.A. = 0.28) to an area of about 80 $\mu$m$^2$ on the sample. The wave vector of the pump is along the crystal $c$-axis (the surface normal), i.e. the electric field vector $\vec{E}$ is orthogonal to the $c$-axis ($\vec{E} \perp c$). The maximum deviation from normal incidence is 8$^{\circ}$ in air and $\sim 2.5 ^{\circ}$ in the crystal for this objective. The polarization and pump flux ($P$) of the pump laser is controlled by liquid-crystal-based devices without mechanical moving parts. The samples are excited with pump flux $P$ from $0.1$ $P_0$ to $P_0$, where $P_0=2.6\times10^{14}$~cm$^{-2}$ photons per pulse. We estimate the photoexcitation density to be from $\approx 2\times10^{16}$ cm$^{-3}$ to $3.4\times10^{17}$ cm$^{-3}$ ($2.7\times10^{-10}$ cm$^{-2}$ per layer) considering the absorption coefficient at 2.1 eV ($\approx10^3$ cm$^{-1}$ for $\vec{E} \perp c$) and Fresnel loss from reflection. The photoexcited carrier density is below the Mott transition \cite{pavesi1989} of direct excitons occurring near e-h pair densities of $4\times10^{17}$ cm$^{-3}$.

Time- and polarization-resolved PL measurements allow us to separately determine the recombination time, the initial spin orientation, and the spin relaxation time. Polarized PL measurements are performed under excitation with excess energy about 0.1 to 0.2 eV above the exciton emission peak. The band-edge exciton PL emission at room temperature is near 620 nm (2.0 eV), independent of thickness \footnote{The exciton peak gradually red shifts from 590 nm (2.1 eV) to 620 nm (2.0 eV) in bulk ($> 1000$ nm) to 90-nm nanoslabs at T = 10 K. This shift is attributed to increasing contribution to PL from localized excitons, which is the subject of ongoing studies and beyond the scope of this paper.}. Additionally, we observe that the quantum yield of luminescence is greatly suppressed in sub-50-nm thick samples \cite{tang2014}. In this paper, we focus on GaSe nanoslabs ranging from about 90 to 2000 nm thick. 

\begin{figure}[htbp]
\includegraphics[width=0.48 \textwidth]{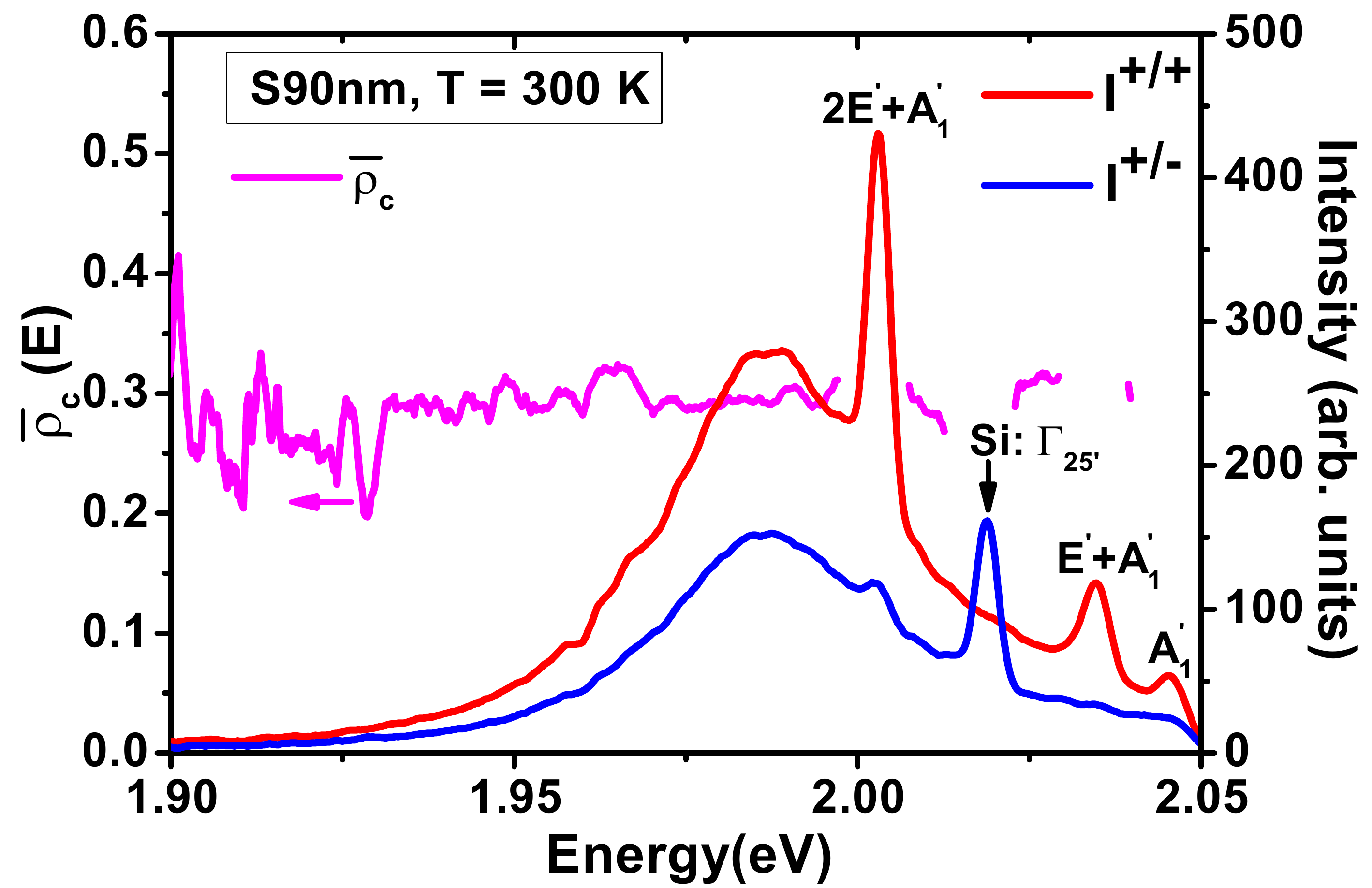}
\caption{\label{fig:PL300K}(color online)  \textbf{Polarized time-integrated PL spectra at room temperature.} Time-integrated PL spectra [$I^{+/+}(E)$(co-circular excitation and detection, red) and $I^{+/-}(E)$(cross-circular excitation and detection, blue)] and degree of circular polarization $\bar{\rho}_c(E)=\frac{I^{+/+}(E)-I^{+/-}(E)}{I^{+/+}(E)+I^{+/-}(E)}$ (magenta) of a 90 nm-thick sample (S90nm) under $\sigma^+$ excitation with 594 nm (2.087 eV) excitation and pump flux $P$ = 0.5 $P_0$ at T = 300 K. Spectra under $\sigma^-$ excitation are symmetric with respect to those under $\sigma^+$ excitation (not shown). The labeled spectral peaks are one-phonon Raman mode $A'_{1}$ (310 cm$^{-1}$) and multi-phonon Raman modes $E'(LO)+A'_1$ [255 + 136 cm$^{-1}$] and $2E'(LO)+A'_1$ [510 + 136 cm$^{-1}$] in GaSe. The multi-phonon Raman modes ($\sim$394 and 648 cm$^{-1}$) become dominant in sub-100-nm thick GaSe and have not been reported in bulk GaSe. Raman shifts are calibrated against the 520-cm$^{-1}$ Raman line ($\Gamma_{25'}$ phonon mode) of the silicon substrate. 
}
\end{figure}

\begin{figure}[htbp]
\includegraphics[width=0.48 \textwidth]{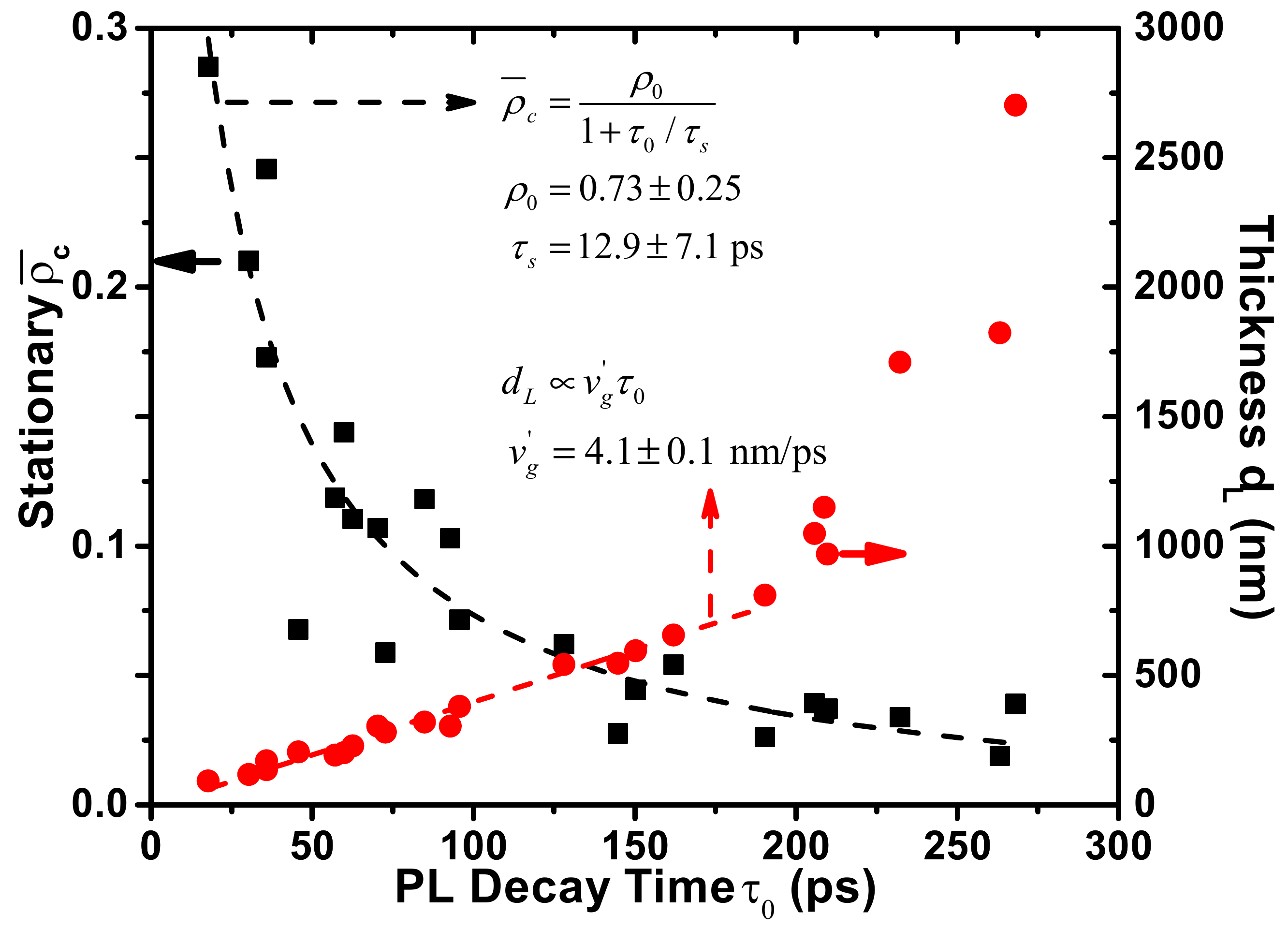}
\caption{\label{fig:TRPL300K}(color online) \textbf{PL dynamics and polarization properties versus sample thickness at room temperature.} PL decay time constant ($\tau_0$) as a function of thickness ($d_L$) of the samples (red circles). $\tau_0$ decreases linearly with thickness for $d_L\lesssim$ 700 nm yielding a slope or an effective group velocity of 4.1 nm/ps. This results in an increasing stationary (time-averaged) circular polarization (black squares) near the exciton PL peak [$\bar{\rho}_c(E\approx1.98 \, \text{eV})$] with decreasing $\tau_0$ (thickness).
}
\end{figure}

\begin{figure}[htbp]
\includegraphics[width=0.48 \textwidth]{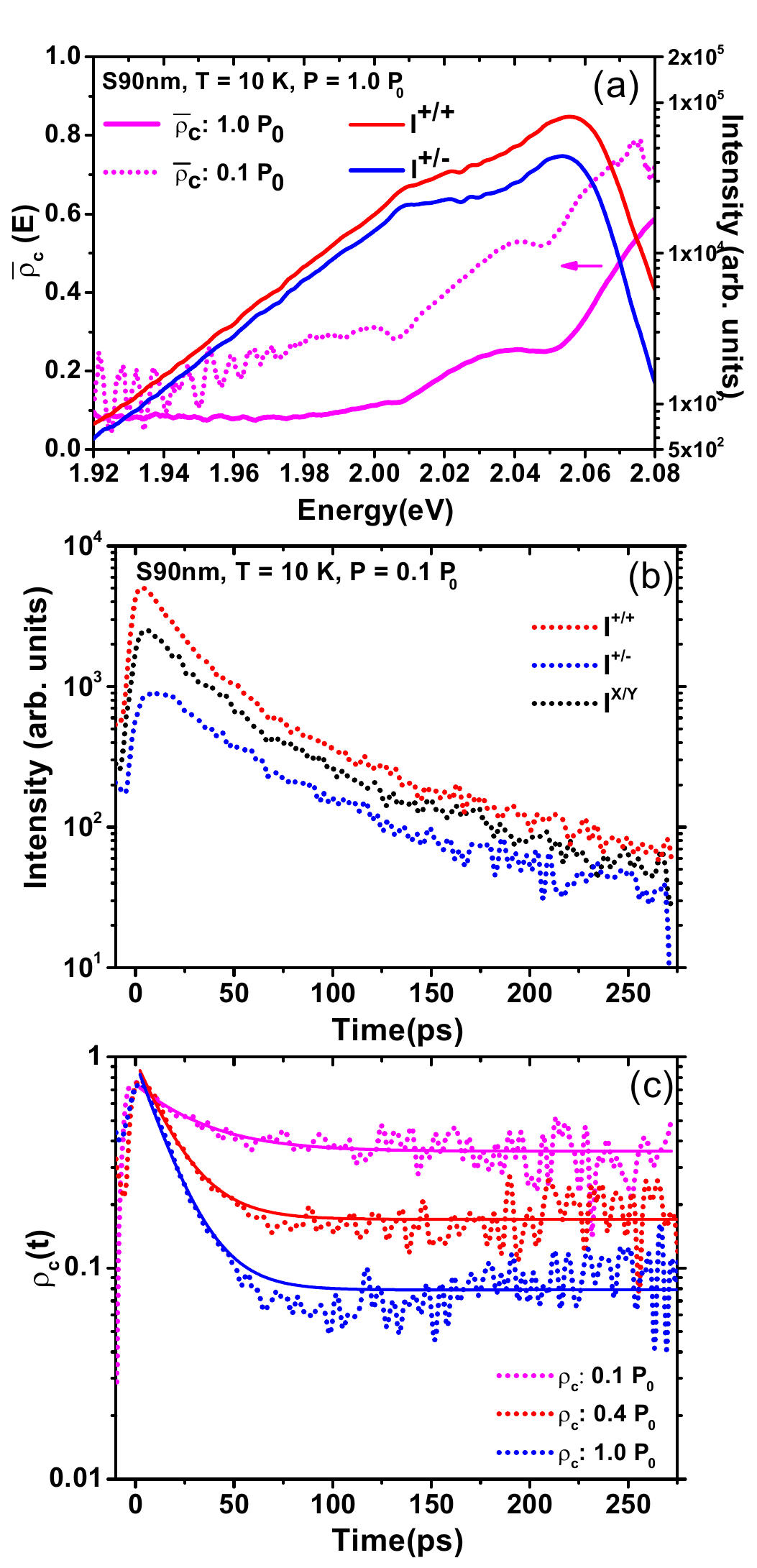}
\caption{\label{fig:PL10K}(color online) \textbf{Polarized PL spectra and dynamics at T = 10 K.} (a) Time-integrated PL spectra [$I^{+/+}(E)$(co-circular, red) and $I^{+/-}(E)$(cross-circular, blue)] and degree of circular polarization $\bar{\rho}_c(E)$ of S90nm under $\sigma^+$ excitation at pump flux $P$ = 1.0 $P_0$. $\bar{\rho}_c(E)$ at $P$ = 0.1 $P_0$ for S90nm is shown as magenta dotted curve for comparison. (b) Polarized time-dependent PL [$I^{+/+}(t)$, $I^{+/-}(t)$, and $I^{X/Y}(t)$ (dashed red, blue, and black, respectively)] of S90nm under $\sigma^+$ or $\sigma^X$ excitation at $P$ = 0.1 $P_0$. (c) $\rho_c(t)$ of PL from S90nm under $\sigma^+$ excitation at a pump flux of $P$ = 0.1, 0.4, and 1.0 $P_0$ (magenta, red, and blue, respectively). Solid lines are bi-exponential fittings.
}
\end{figure}

\paragraph{Results.} Upon absorption of circularly polarized light above the band-gap, the stationary (time-averaged) degree of circular polarization ($\bar{\rho}_c$) of luminescence represents photoexcited carrier spin memory. In Fig. \ref{fig:PL300K}, we show the polarized time-integrated PL spectra [$I(E)$] for a 90 nm-thick sample (S90nm) at room temperature. Across a range of sample thicknesses ($d_L$), time-integrated PL shows a pronounced increase of $\bar{\rho}_c$ from $\sim$0.05 in bulk to 0.3 in sub-100-nm nanoslabs. Corresponding polarized time-dependent PL reveals that the degree of circular polarization $\rho_c(t)$ decays within 10 ps, independent of the slab thickness and photoexcited carrier density \cite{tang2014}. In contrast, the total (unpolarized) PL decays exponentially with a time constant ($\tau_0$) that increases linearly with thickness from about 20 ps to 250 ps for 90 nm $\lesssim d_L \lesssim$ 700 nm (Fig \ref{fig:TRPL300K}). The fast PL rise and decay time constants in sub-100-nm GaSe nanoslabs are comparable to the $\sim$10-ps spin relaxation time, resulting in the preservation of a high spin polarization during the brief absorption-cooling-emission cycle \cite{nusse1997}. In Fig. \ref{fig:TRPL300K}, we also plot $\bar{\rho}_c$ as a function of PL decay time $\tau_0$ for \emph{all} samples studied. The stationary $\bar{\rho}_c$ is well described by a simple model: $\bar{\rho}_c = {\rho_0}/{(1+\tau_0/\tau_s)}$, where $\tau_0$ is the lifetime of photoexcited carriers, and $\tau_s$ is the spin relaxation time. Fitting yields $\rho_0 = 0.73 \pm 0.25$ and $\tau_s=13 \pm 7$ ps, consistent with the time-resolved measurements. Therefore, the apparent enhancement of ``spin memory'' revealed in time-integrated PL is largely due to the linear decrease of the PL decay time ($\tau_0$) with thickness, rather than a long spin relaxation time.

At cryogenic temperature (T = 10 K), the spin and PL decay dynamics with 560 nm (2.214 eV) to 575 nm (2.156 eV) excitation slow markedly. Similar $\bar{\rho}_c(E)$ and PL dynamics are observed for all studied samples with $d_L >$ 90 nm at T = 10K. We thus describe only the results in 90 nm-thick sample S90nm. At a pump fluence of 0.1 $P_0$, the stationary $\bar{\rho}_c(E)$ reaches 0.8 at the high-energy range of the exciton emission spectrum and decreases gradually to about 0.3 at low emission energies (Fig. \ref{fig:PL10K}a). Meanwhile, time-dependent PL decay becomes bi-exponential with time constants $\tau_0' \, \approx$ 20-50 ps and $\tau_0'' \, \approx$ 150-200 ps (Fig. \ref{fig:PL10K}b). In contrast, the PL rise time remains less than $10$ ps (limited by instrument temporal resolution), suggesting sub-10-ps cooling and momentum relaxation of radiative photoexcited carriers independent of temperature. Finally, time-dependent circular polarization $\rho_c(t)$ is also found to be bi-exponential, with decay time constants $\tau_s' \, \approx$ 30--40 ps and $\tau_s'' \, \gtrsim$ 500 ps (Fig. \ref{fig:PL10K}c). We reproduce quantitatively the bi-exponential decays of both PL intensity and polarization using a rate-equation model \cite{vinattieri1994,tang2014}. The shorter time constant $\tau_s'$ shows a weak power-law dependence on the photoexcited density ($n_X$) with $\tau_s' \propto n_X^{(-0.23\pm0.06)}$, while there is no identifiable dependence on thickness for either population decay or spin relaxation.

\begin{figure}[htb]
\centering
\includegraphics[width=0.48 \textwidth]{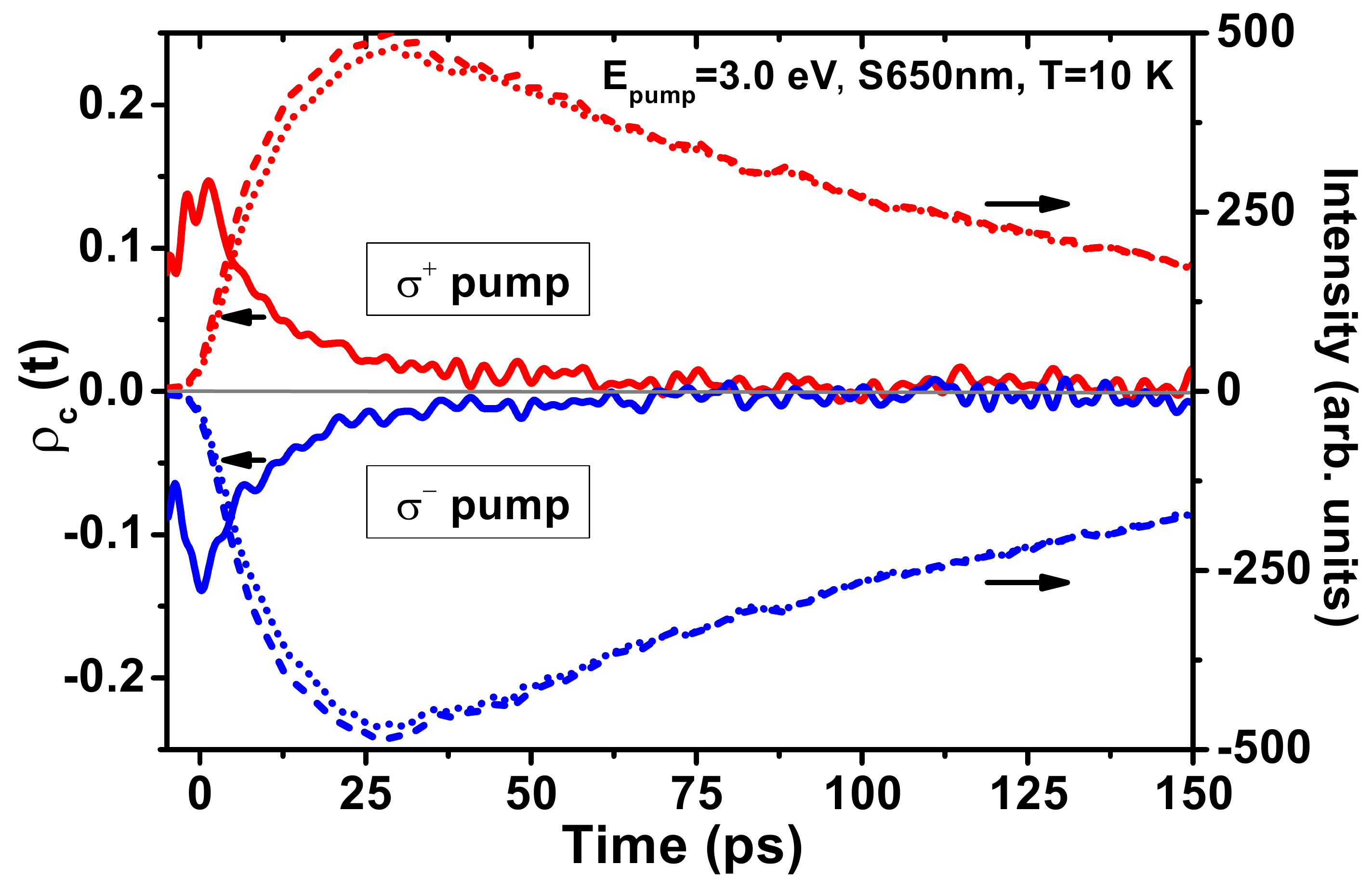}
\caption{\label{fig:ex413nm}(color online) \textbf{Polarized time-dependent PL under optical excitation at 3 eV.} Time-dependent PL intensities $I^{\pm/\pm}(t)$ (co-circular, dashed curves) and $I^{\pm/\mp}(t)$ (cross-circular, dotted curves), degree of circular polarization $\rho_c(t)$ (solid curves) under excitation at $E_{pump}$ = 3 eV and T = 10 K. Red and blue curves are for $\sigma^+$ and $\sigma^-$ excitation, respectively. The decay of $\rho_c(t)$ is slower at T = 10 K than that at T = 300 K (not shown) ($\approx$7 ps vs. 3 ps) as is the PL rise time ($\approx$18 ps vs. 5 ps). 
}
\end{figure}

To explore the limits of spin memory, we also performed experiments on a 650 nm-thick sample (S650nm) under 3.0 eV excitation (Fig.~\ref{fig:ex413nm}). We find that the initial $\rho_0$ at the band edge is 0.15, irrespective of temperature. By contrast, a PL circular polarization of $\sim$0.005--0.009 has been observed in bulk GaAs under a $3.00\pm0.05$ eV excitation (i.e., excess energy $\approx$ 1.5 eV above the GaAs band gap) when the $L$-valley transitions are involved \cite{zhang2013}.  The reduction of $\tau_s'$ with increasing carrier density and $\rho_0$ with increasing excitation energy suggest that the Elliot-Yafet (EY) mechanism plays a larger role than the Dyakonov-Perel (DP) mechanism \cite{dyakonov1984,dyakonov2008,pikus1984,wu2010} in the initial spin relaxation for non-thermal carriers at low temperature. 

\paragraph{Discussion.} The generation of a substantial degree of spin polarization at the band edge after highly nonresonant excitation is a consequence of (1) a large degree of spin polarization of the initially created high-energy carriers and (2) spin relaxation rates that are slow compared to energy and momentum relaxation as well as the band-edge PL decay rates. All samples display a nearly instantaneous rise ($\sim$5 ps, resolution-limited) in the band-edge PL under excitation at excess energies $\lesssim$300 meV at room and cryogenic temperatures, indicating sub-10-ps momentum and energy relaxation of photoexcited carriers toward $k$ =0 ($\Gamma$ point). This time scale is consistent with the momentum scattering time of 5 ps deduced through steady-state measurements \cite{gamarts1977,ivchenko1977}. Carriers thus spend limited time in high-momentum states where spin relaxation is fastest. In contrast, at 3.0 eV excitation, carriers take markedly longer to dissipate 1 eV excess energy, especially at T = 10 K. This increase of energy relaxation time partially accounts for the reduced initial degree of spin polarization observed under 3.0 eV excitation. 

The decrease of PL decay $\tau_0$ with thickness for $d_L<$ 700 nm at $T=300$ K means that a greater fraction of the emission occurs before the spin polarization decays, leading to larger $\bar{\rho}_c$ in sub-100-nm GaSe nanoslabs. The decrease of $\tau_0$ is unlikely due to an enhancement of radiative recombination rates with dimensionality and confinement \cite{andreani1991} given that the bulk Bohr exciton radius of GaSe is only $\sim$5 nm. The linear dependence of $\tau_0 \propto d_L$ can result from the propagation of exciton-polaritons \cite{aaviksoo1991} or non-radiative surface recombination \cite{aspnes1983} as described in Ref. \cite{tang2014}. The PL measurements presented here only probe excitons and carriers near the $\Gamma$ point. Nevertheless, the distinct PL dynamics at room and cryogenic temperatures suggest that there is significant non-radiative recombination due to inter-valley scattering between the $\Gamma$ and M valleys \cite{capozzi1993}. 

The reduced mixing of valence bands makes GaSe and other group-III monochalcogenides such as GaS and InSe \cite{kuroda1980,do2015} materials with potential for devices relying on the ability to generate and maintain high degrees of spin polarization. In these materials, the hole spin relaxation rate is expected to decrease significantly compared to that in III-V semiconductors because the EY relaxation mechanism is suppressed by the reduced mixing of distant valence bands. This hypothesis is supported by the quantitative exciton and spin relaxation rates obtained by fitting the experimental polarized PL dynamics as described in Ref. \cite{tang2014}. Nanoscale multi-layer GaSe or InSe are, in principle, less challenging to synthesize and integrate with micro- and nano-fabrication processes compared to atomic membranes of TMDs. Additionally, non-centrosymmetric group-III monochalcogenides allow for more direct optical control of light-matter coupling beyond the monolayer regime.

\begin{acknowledgements}
	This work is supported by Michigan State University and by NSF through DMR-09055944. This research has used the W. M. Keck Microfabrication Facility. We thank Brage Golding, Bhanu Mahanti, and Carlo Piermarocchi for comments and discussions.
\end{acknowledgements}


%

\end{document}